\documentclass{aastex}
\usepackage{graphics,emulateapj5,onecolfloat}

\newcommand{\lsim}{\raisebox{-0.3ex}{\mbox{$\stackrel{<}{_\sim} \,$}}}

\begin{document}
\twocolumn[

\shorttitle{Point sources and SZ surveys}
\shortauthors{White and Majumdar}

\title{Point Sources in the Context of Future SZ Surveys}
\author{Martin White \altaffilmark{1} and Subhabrata Majumdar\altaffilmark{2}}
\affil{${}^1$Departments of Physics and Astronomy, University of
California, Berkeley, CA 94720}
\affil{${}^2$CITA, University of Toronto, 60 St George St, Toronto, ON,
M5S3H8}
\email{mwhite@astron.berkeley.edu}\email{subha@cita.utoronto.ca}

\begin{abstract}
We look at the impact of Infra-Red (IR) and radio point sources on
upcoming large yield Sunyaev-Zel'dovich (SZ) cluster surveys such as
APEX, SPT and ACT. The IR and radio point source counts are based on
observations by the the SCUBA and WMAP instruments respectively.
We show that the contributions from IR source counts, when extrapolated
{}from the SCUBA frequency of $350\,$GHz to the operating frequencies of
these surveys ($\sim 100-300\,$GHz), can be a significant source of
additional `noise' which needs to be accounted for in order to
extract the optimal science from these surveys.
Alternatively, these surveys give us an opportunity to study IR sources,
their numbers and clustering properties, opening a new window to the
high redshift universe.
For the radio point sources, the contribution depends on a more uncertain
extrapolation from $40\,$GHz but is comparable to the IR near $200\,$GHz.
However, the radio signal may be correlated with clusters of galaxies
and have a disproportionately larger effect.
\end{abstract}								       
									
]

\section{Introduction} \label{sec:intro}

The study of anisotropies in the Cosmic Microwave Background (CMB) has
proven to be a gold-mine for cosmology.  The primary anisotropies on scales
larger than $10'$ have now been probed with high fidelity by WMAP
(Bennett et al.~\cite{WMAP}) over the whole sky, leading to strong
constraints on our cosmological model. Within the next few years this
activity will be complemented by high angular resolution, high sensitivity
observations of secondary anisotropies by the
SZA\footnote{http://astro.uchicago.edu/sza/},
APEX-SZ experiment\footnote{http://bolo.berkeley.edu/apexsz/},
the South Pole Telescope (SPT\footnote{http://astro.uchicago.edu/spt/})
and the Atacama Cosmology Telescope
(ACT\footnote{http://www.hep.upenn.edu/$\sim$angelica/act/act.html})
which are aiming to make arcminute resolution maps with $10 \mu$K sensitivity
(or better) at millimeter wavelengths.

The dominant secondary anisotropy is expected to be the Compton scattering of
cold CMB photons from hot gas along the line of sight, known as the thermal
Sunyaev-Zel'dovich (SZ) effect (Sunyaev \& Zel'dovich \cite{SZ72,SZ80};
for recent reviews see Rephaeli \cite{Rep} and Birkinshaw \cite{Bir})
though other signals can be present at lower amplitudes.
In principle, a measurement of such anisotropies would constrain cosmological
parameters (Weller et al.~\cite{weller02}, Levin et al.~\cite{levin02},
Majumdar \& Mohr \cite{majumdar03}, Hu \cite{hu03b}),
probe the thermal  history of the intra-cluster medium
(Majumdar \cite{majumdar01}, Zhang \& Pen \cite{zhang02}),
put useful constraints on re-ionization models through the kinetic
SZ effect (Zhang et al.~\cite{zhang03}) and allow us to map the dark
matter back to the surface of last scattering
(Seljak \& Zaldarriaga \cite{seljak99};
 Zaldarriaga \& Seljak \cite{zaldarriaga99};
 Hu \cite{hu01}; Hirata \& Seljak \cite{hirata03};
 Okamoto \& Hu \cite{okamoto03}).

However, like the primary anisotropies, the secondary signals must be
disentangled from other astrophysical emissions at the observed frequencies.
One of the major source of contamination on small scales are point sources
(by which we shall mean throughout sources unresolved at arcminute resolution).
The purpose of this paper is to outline how point sources impact high
resolution CMB experiments and the associated `confusion' noise, describe how
one can make small extrapolations of existing data to estimate the amplitude
of this noise and finally discuss the science which can come from studies of
this source population.
We make use of new observational constraints on the number of sources at
frequencies and flux level relevant to APEX-SZ, SPT and ACT.

\section{Point sources as noise} \label{sec:theory}

We will see that point sources will be particular troublesome for high
resolution surveys, and we will attempt to quantify their effect in terms of
an equivalent ``noise''.  Such a quantification can only be approximate,
however it is a useful metric by which to judge the relative importance of
different components and to plan survey strategies. Physically, the IR point
sources are thought to be dusty, high redshift galaxies and so we do not
expect them to be correlated with any particular place on the sky or any
particular low redshift source (such as a cluster of galaxies). We note,
however, that the counts are sufficiently steep that lensing by the larger
clusters can non-trivially increase the number of sources observed above a
given flux cut and this should be taken into account in interpreting our
results (see e.g.~Perrotta et al.~\cite{PMBBZGSD} for recent theoretical
modelling). The radio sources are more problematic (Holder \cite{holder02}),
as they are likely to be correlated with the SZ signal from clusters of
galaxies to some extent.
Since this correlation is difficult to quantify at present we
will simply provide ``noise'' estimates for them as well.

To proceed, we calculate the angular power spectrum for a population of
point sources described by a flux distribution and a clustering amplitude.
If we assume that the number of sources of a given flux is independent
of the number at a different flux, and if the angular two-point function
of the point-sources is $w(\theta)$, then the angular power spectrum,
$C_\ell$, contributed by these sources is (Scott \& White \cite{ScoWhi})
\begin{equation}
  C_\ell(\nu)=\int_0^{S_{\rm cut}}\ S_\nu^2\
  {{\rm d}N\over {\rm d}S_\nu} \,dS_\nu
  + w_\ell \left(I_{\nu}\right)^2\!,
\label{eqn:cltot}
\end{equation}
assuming that all sources with $S{>}S_{\rm cut}$ are removed.
Here, $I_{\nu}=\int S\, {\rm d}N/{\rm d}S\, dS$ is the background
contributed by sources below $S_{\rm cut}$.
Following the conventional notation, $C_\ell$ is the Legendre transform
of the correlation function $C(\theta)$ produced by the sources and
$w_\ell$ is the Legendre transform of $w(\theta)$:
\begin{eqnarray}
C(\theta) & = & {\displaystyle 1\over 4\pi} \sum_{\ell} (2\ell+1)C_\ell
  P_\ell(\cos\theta) \\
w(\theta) & = & {\displaystyle 1\over 4\pi} \sum_{\ell} (2\ell+1)w_\ell
  P_\ell(\cos\theta),
\end{eqnarray}
with $P_\ell(\cos\theta)$ the Legendre polynomial of order $\ell$.
The first term in equation~(\ref{eqn:cltot}) is the usual Poisson shot-noise
term (see Peebles \cite{LSSU} \S46, or Tegmark \& Efstathiou \cite{TegEfs}),
the second is due to clustering, assuming that the clustering is independent
of flux.
All reasonable $dN/dS$ give a $C_\ell$ which converges at the faint end and
is very insensitive to the upper flux density cut.

Because these sources are contributing a foreground to CMB anisotropy
experiments we will recast our results in `temperature' units. 
This is easily done by applying the conversion factor between flux and
temperature which is given by
\begin{eqnarray}
{\partial B_\nu\over\partial T} &=&
  {2k\over c^2} \left( {kT_{\rm CMB}\over h} \right)^2
  {x^4 {\rm e}^x\over ({\rm e}^x-1)^2} \nonumber\\
&  = & \left( {99.27\,{\rm Jy}\ {\rm sr}^{-1}\over \mu\,{\rm K}} \right)
  {x^4 {\rm e}^x\over ({\rm e}^x-1)^2},
\end{eqnarray}
where $B_\nu$ is the Planck function, $k$ is Boltzmann's constant,
$x\equiv h\nu/k_BT_{\rm CMB}=\nu/56.84\,$GHz is the
`dimensionless frequency' and
$1\,{\rm Jy}\,{=}\,10^{-26}\,{\rm W}\,{\rm m}^{-2}{\rm Hz}^{-1}$.
Conveniently a $10\,\mu$K CMB fluctuation in a $1'$ FWHM pixel gives a
flux close to $1$mJy viz:
\begin{eqnarray}
  S_\nu &=& 0.3\,{\rm mJy} \left({\sigma\over 10\mu{\rm K}}\right)
     \left({\theta\over 1'}\right)^2 \qquad @150{\rm GHz} \\
  S_\nu &=& 0.4\,{\rm mJy} \left({\sigma\over 10\mu{\rm K}}\right)
     \left({\theta\over 1'}\right)^2 \qquad @220{\rm GHz} \\
  S_\nu &=& 0.3\,{\rm mJy} \left({\sigma\over 10\mu{\rm K}}\right)
     \left({\theta\over 1'}\right)^2 \qquad @350{\rm GHz}
\end{eqnarray}
indicating that these upcoming surveys will be probing the source
population at the mJy level.

Since the clustering of the sources is highly uncertain\footnote{Though we
expect this to improve soon with the release of data from the SHADES survey,
http://www.roe.ac.uk/ifa/shades/, and of course APEX-SZ, SPT and ACT.}
at present, we will set $w_\ell=0$, though we shall return to this issue in
\S\ref{sec:firb}.
For now this is a conservative assumption as we expect the sources to be
non-trivially clustered if they are associated with rare, highly biased
tracers of the density field at high-$z$ but we also expect those correlations
to be most important at low-$\ell$.
In the Poisson limit the $C_\ell$ are independent of $\ell$, i.e.~they have
the same shape as Poisson/white noise.
We therefore express our results in terms of an `effective' point source
noise $\sigma_{\rm pix}$ per pixel of FWHM $\theta_{\rm pix}$ using
\begin{equation} 
  C_\ell = \left(\theta_{\rm pix}\sigma_{\rm pix}\right)^2 \,,
\label{eqn:spix}
\end{equation} 
and express $\sigma_{\rm pix}$ in $\mu$K for a fiducial $1'$ pixel.
This value can be rescaled to any desired resolution using Eq.~\ref{eqn:spix}.
This $\sigma_{\rm pix}$ is closely related to the standard `confusion noise'
often quoted by radio astronomers.  To make the connection more explicit let
us write 
\begin{equation} 
  {dN\over dS} \equiv {N_0\over S_0}\ g\left({S\over S_0}\right) 
\end{equation} 
which defines $g(x)$.  Then the `noise' induced by the sources is
$\sigma_{\rm conf}=S_0\sqrt{N_{\rm pix}}\,{\cal I}(x)$ with
$x=S_{\rm cut}/S_0$, $N_{\rm pix}=N_0\theta_{\rm pix}^2$ and 
\begin{equation} 
  {\cal I}^2(x) = \int_0^x\ ds\, s^2g(s) \qquad .
\end{equation} 
If we assume we can subtract sources brighter than $n\sigma$
this simplifies to 
\begin{equation}
  \sigma_{\rm conf} = S_0\left({x\over n}\right)
  \quad {\rm where}\quad
  {x\over {\cal I}(x)} = n\sqrt{N_{\rm pix}} \ . 
\label{eqn:sigconf} 
\end{equation} 
Upon converting between flux and temperature units and accounting for the
flux cut this expression is the same as Eq.~\ref{eqn:spix}.
We mention finally one additional complication which we can include.
If the sources have a frequency spectral index distribution ${\cal F}(\beta)$
and we assume different sub-populations are independent then we can
generalize the above to 
\begin{equation}
  \sigma_{\rm conf}^2 = N_{\rm pix}\int d\beta\ {\cal F}(\beta)
  S_0^2(\beta) {\cal I}^2\left(x(\beta)\right)
\end{equation}
where $S_0(\beta)$ indicates the value of $S_0$ obtained by extrapolating
the fiducial $S_0$ to the required frequency using $S_\nu\propto\nu^\beta$.
This tends to be a relatively small effect.  Even a 30\% gaussian uncertainty
in $\beta$ with $\langle\beta\rangle=0$ (see \S\ref{sec:counts} below)
increases the confusion noise at $5\,$mJy by only 10\% over the
$\beta\equiv 0$ value.  For this reason we shall work with uniform,
constant $\beta$ from now on.

\section{Noise Estimates} \label{sec:counts}

The remaining step in computing the effective noise contributed by point
sources is thus a model for the counts.  While it is theoretically quite
difficult to predict these functions, we are fortunate to have observations
at close to the relevant frequencies and flux levels.
It is thus more robust to extrapolate the observations than to start from
an {\it ab initio\/} theoretical model.

\subsection{Radio sources}

For the radio point sources, a fit to the Q-band data from WMAP
(Bennett et al.~ \cite{WMAP}) can be written as
\begin{equation}
  {dN\over dS_\nu} = {N_0\over S_0} \left( {S_\nu\over S_0} \right)^{-2.7}
\end{equation}
where $N_0=1200\,{\rm deg}^{-2}=4\times 10^6\,{\rm sr}^{-1}$ with an
uncertainty of around 30\% and $S_0\simeq 1\,$mJy.
In obtaining these numbers we need to extrapolate from very high flux
levels to mJy levels, and this extrapolation is sensitive to the slope of
the distribution.  There is evidence that the slope flattens from -2.7 to
-2.2 at lower fluxes.  We shall choose -2.3 as a compromise, since
$S^{5/2}dN/dS$ is roughly flat in Fig.~13 of Bennett et al.~(\cite{WMAP}).
For $S_0=1\,$mJy this lowers $N_0$ to $80\,{\rm deg}^{-2}$.
The extrapolation to higher frequency is also somewhat uncertain.
The primary emission mechanism in these sources is synchrotron emission,
which for a power-law electron spectrum would give a power-law $S_\nu$.
However synchrotron `aging' and optical depth effects can give departures
{}from a power-law spectrum, allowing a wide range of spectral shapes
including spectra which peak at or above $100\,$GHz.
Advection dominated accretion onto a super-massive black hole is also
characterized by a strongly inverted spectrum peaking at millimeter
wavelengths, as are radio afterglows of gamma-ray bursts.
In addition, many of the sources with significant flux at high frequency are
thought to be young, compact sources which are likely to be highly time
variable.

Given these large uncertainties, we will attempt to bracket the reasonable
range and extrapolate the fluxes to higher frequency assuming two different
power-law spectral indices, $S_\nu\propto\nu^\beta$ with $\beta=-0.3$
(Tegmark \& Efstathiou \cite{TegEfs}) and $\beta=0$
(close to the mean of the distribution in Trushkin \cite{Tru}).
At the mJy level we have close to one radio source per 50 beams if the point
sources predominantly have a flat (or rising) spectrum.

\subsection{IR sources}

\begin{figure}
\begin{center}
\resizebox{3.3in}{!}{\includegraphics{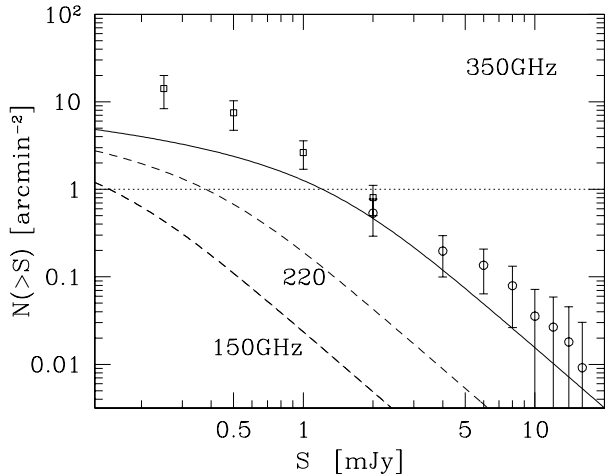}}
\end{center}
\caption{The cumulative SCUBA source counts at $350\,$GHz as a function of
flux, $S$, from Smail et al.~\cite{SIB} (squares) and
Borys et al.~\cite{borys99,borys03} (circles).
The fit of Borys et al.~\cite{borys03} is shown as the solid line, and the
extrapolations to $220\,$GHz and $150\,$GHz assuming $S_\nu\propto\nu^{2.5}$
as dashed lines.
Upcoming surveys will be sensitive to sources with $S\sim 1\,$mJy.}
\label{fig:counts}
\end{figure}

\begin{figure}
\begin{center}
\resizebox{3.3in}{!}{\includegraphics{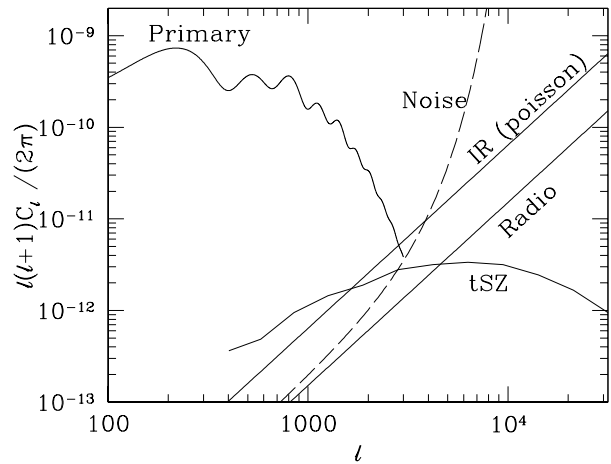}}
\end{center}
\caption{The angular power spectra of various sources at $150\,$GHz.
The line labeled ``primary'' is the primary CMB anisotropy spectrum,
the line labeled ``tSZ'' is the thermal SZ spectrum from
White, Hernquist \& Springel \protect\cite{WHS} and is uncertain at the
factor of 2 level.  The (dashed) line labeled ``noise'' is the instrument
noise assuming $10\,\mu$K per $1'$ pixel and a resolution of $1'$.
The two unlabeled lines rising rapidly to the top right of the plot are
the Poisson contribution of the radio and IR sources for $S_{\rm cut}=5\,$mJy,
assuming $\beta=0$ and $2.5$ respectively.}
\label{fig:lcl150}
\end{figure}

For the IR sources we can use observations at $350\,$GHz made with the
Submillimeter Common-User Bolometer Array
(SCUBA; Holland et al.~\cite{Holland})
on the James Clerk Maxwell Telescope. SCUBA has been used to make several
deep observations
(Barger et al.~\cite{Barger}; Eales et al.~\cite{Eales};
 Holland et al.~\cite{Holletal}; Hughes et al.~\cite{Hughes};
 Smail et al.~\cite{SIB})
{}from which we can extract source counts.
In particular we shall use the recent work of Borys et al.~\cite{borys03}
who give a phenomenological fit to the counts at $350\,$GHz:
\begin{equation}
  {dN\over dS_\nu} =
    {N_0 \over S_0} \left[\left(S_\nu\over S_0\right)
     +  \left(S_\nu\over S_0\right)^{3.3} \right]^{-1}
\label{eqn:IRcount}
\end{equation}
with $N_0=1.5\times 10^4\,{\rm deg}^{-2}=4.9\times 10^7\,{\rm sr}^{-1}$ and
$S_0\simeq 1.8\,$mJy.
This model provides a reasonable fit to the existing data near $1\,$mJy
that does not over-produce the far-infrared background (FIB) light
(Puget et al.~\cite{puget96}).
We estimate that the uncertainty in the normalization is roughly a factor
of 2, due to the small sky area surveyed.
We will extrapolate from $350\,$GHz to lower frequency using
$S_\nu\propto\nu^\beta$ with $\beta=2.5$, close to the typical spectrum
of a dusty star-bust galaxy at high-$z$.
We will also examine the effect of steeping the frequency dependence to a
more conservative value $\beta=3$.

In figure \ref{fig:counts}, we have plotted the observed IR source counts
of (Borys et al.~\cite{borys99}, Smail et al.~\cite{SIB}) at $350\,$GHz
along with our extrapolation to lower frequencies.
Note that at $150\,$GHz there would be more than 100 IR point sources above
$1\,$mJy per deg${}^2$!
This means tens to hundreds of thousands of IR sources (depending on the
survey).  These sources need to be carefully accounted for when considering
secondary CMB science.

\begin{figure}
\begin{center}
\resizebox{3.3in}{!}{\includegraphics{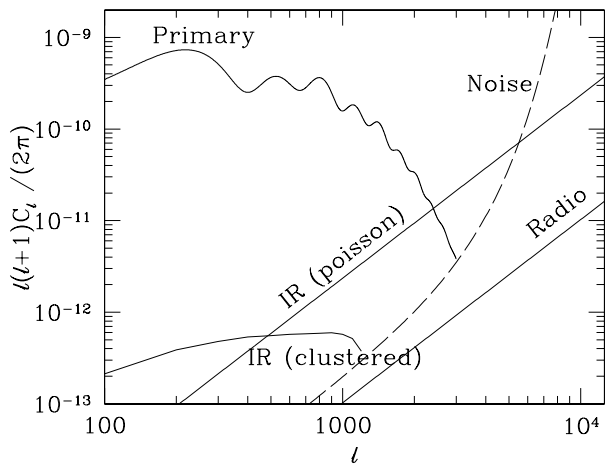}}
\end{center}
\caption{The angular power spectra of various sources at $220\,$GHz,
the thermal SZ ``null''.  Lines are as in Fig.~\protect\ref{fig:lcl150}
except that a model of the correlations in the FIB from
Knox et al.~\protect\cite{knox01} is also shown.}
\label{fig:lcl220}
\end{figure}

\section{Results} \label{sec:results}

\begin{table}
\begin{center}
\begin{tabular}{ccc|cc}
$S_{\rm cut}$ (mJy) & \multicolumn{4}{c}{$\sigma_{\rm pix}$($\mu$K)} \\
\hline
        & \multicolumn{2}{c|}{$\beta=0$} & \multicolumn{2}{c}{$\beta=-0.3$} \\
        &    $150\,$GHz  & $220\,$GHz    &    $150\,$GHz  & $220\,$GHz \\
     1  &   5   &   4  &  4 &   3 \\
     5  &   9   &   8  &  7 &   5 \\
    10  &  12   &  10  &  9 &   7 \\
    50  &  20   &  17  & 16 &  12 
\end{tabular}
\end{center}
\caption{The effective noise for Poisson distributed radio sources at $150$
and $220\,$GHz assuming a fiducial $1'$ pixel.  We show extrapolations from
$40\,$GHz assuming a spectral index of $\beta=0$ and $-0.3$.}
\label{tab:radio}
\end{table}

\begin{table}
\begin{center}
\begin{tabular}{ccc|cc}
$S_{\rm cut}$ (mJy) & \multicolumn{4}{c}{$\sigma_{\rm pix}$($\mu$K)} \\
\hline
        & \multicolumn{2}{c|}{$\beta=2.5$} & \multicolumn{2}{c}{$\beta=3$} \\
            &    $150\,$GHz  & $220\,$GHz    &    $150\,$GHz  & $220\,$GHz \\
     1  &  15   &  23  & 11 &  20 \\
     5  &  19   &  36  & 13 &  29 \\
    10  &  20   &  39  & 13 &  32 \\
    50  &  21   &  44  & 14 &  35
\end{tabular}
\end{center}
\caption{The effective noise for Poisson distributed IR sources at $150$
and $220\,$GHz assuming a fiducial $1'$ pixel.  We show extrapolations from
$350\,$GHz assuming a spectral index of $\beta=2.5$ and $3$.  Any correlations
in the sources will (likely) increase $\sigma_{\rm pix}$.}
\label{tab:ir}
\end{table}

For $150$ and $220\,$GHz the point source contributions, assuming $1'$ pixels,
are summarized in Tables \ref{tab:radio} and \ref{tab:ir}.
We do not include the effects of source clustering, which would increase these
numbers, or gravitational lensing of the population by a foreground object
such as a cluster.
The corresponding power spectra are shown in Figs.~\ref{fig:lcl150}
(at $150\,$GHz)  and \ref{fig:lcl220} (at $220\,$GHz).
We caution the reader that several of the processes on these figures are
non-Gaussian, so the power spectra tell only some of the story.
Nonetheless we can see that point sources will be a large contribution to
the signal at these frequencies, which will need to be properly accounted
for in any analysis aimed at recovering the thermal SZ effect.
Even more care will be required for processes with lower signal levels or
methods which assume that the underlying map is predominantly Gaussian in
nature (as for example the primary CMB is) since residual emission needs
to be carefully controlled.

\begin{figure}
\begin{center}
\resizebox{3.3in}{!}{\includegraphics{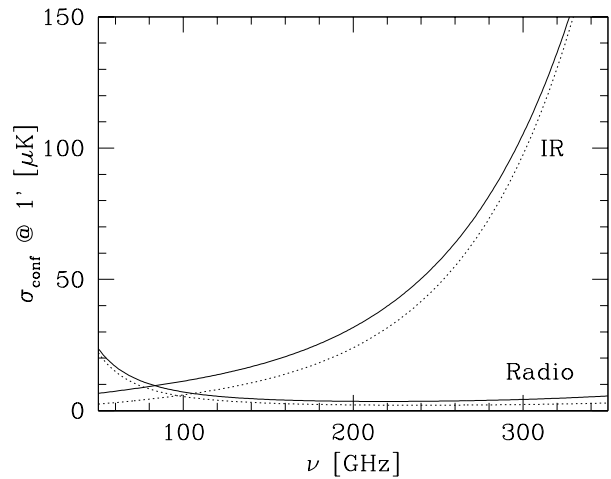}}
\end{center}
\caption{Confusion noise, in $\mu$K, for a $1'$ beam as a
function of frequency.  Contributions from Poisson distributed radio and
IR sources are shown.  The solid lines represent $\beta=0$ and $2.5$
while the dotted lines represent $\beta=-0.3$ and $3$ respectively.}
\label{fig:freq}
\end{figure}

There is one obvious strategy for controlling the effect of these point
sources, and that is to use multiple observing frequencies and multiple
instruments with matched sensitivities and resolutions.

Finally we make an estimate of the `confusion noise' as a function of beam
size at these frequencies.  For the radio sources
${\cal I}(x)=\sqrt{x^{3-2.3}/(3-2.3)}$ allowing us to estimate
$\sigma_{\rm conf}$ as a function of beam size.
If we assume $\beta=0$ the typical values for a $4\sigma$ cut at $150\,$GHz
are $\sigma_{\rm conf}=2\,$mJy at $5'$,
$0.4\,$mJy at $2'$, $0.1\,$mJy at $1'$ and $0.05\,$mJy at $0.5'$.
For the IR sources ${\cal I}(x)$ is a hypergeometric function.
As $x\to 0$, $x/{\cal I}(x)\to\sqrt{2}$ while $x/{\cal I}(x)\propto x$ for
$x\gg 1$.  Assuming $\beta=2.5$ the typical values for a $4\sigma$ cut at
$150\,$GHz are $\sigma_{\rm conf}=3.8\,$mJy at $5'$,
$1.4\,$mJy at $2'$, $0.6\,$mJy at $1'$ and $0.3\,$mJy at $0.5'$.
In thermodynamic units this is e.g.~$20\,\mu$K at $1'$
in agreement with Table \ref{tab:ir}.
We show the frequency dependence of the confusion noise in Fig.~\ref{fig:freq}.

\section{Doing point source studies with APEX/SPT} \label{sec:firb}

The point sources described above represent an analysis challenge for studies
of secondary CMB anisotropies such as the thermal and kinetic SZ effects or
gravitational lensing.  However, they also represent increased science reach
for the surveys in different fields of research.

At present, the SCUBA sources account for 40\% of the $350\,$GHz sub-mm
background (Borys et al.~\cite{borys03}) and thus go a long way towards
resolving the high-$z$ part of the cosmic far-infrared background (FIB).
This suggests that APEX-SZ, SPT and ACT with their high sensitivity and
large areal coverage could become ideal instruments for studying the FIB 
at the faint but especially the bright end of the source counts.
As is evident from Figure \ref{fig:counts}, these future surveys would also
be able to detect thousands of high redshift sources.

These upcoming surveys would be able to probe the FIB correlations which  has
been shown (Haiman \& Knox \cite{haiman00}, Knox et al.~\cite{knox01}) to
give an additional handle on early structure formation at $z>1$. As we show
in Fig.~\ref{fig:lcl220}, detailed observations of the FIB correlations are
possible if we can accurately subtract the contribution from primary
anisotropies and measure power on angular scales of a few degrees.  
Under assumptions similar to those in Knox et al.~\cite{knox01}, we expect
that a $10^2$ square degree survey to around $20\,\mu$K would make precision
measurements (at the several percent level near $\ell\sim 10^3$) of the
clustering of the FIB sources.
Multiple frequency information will be crucial in separating the signal from
other contaminants.  In addition, since the FIB is composed of contributions
from sources at different redshifts, the FIB sky maps at different
frequencies are not perfectly correlated.
The shape of the FIB power spectrum at different frequencies and the
correlations between them can give information about the contributing
sources (Knox et al.~\cite{knox01}).
Studies of the FIB with these surveys would be complimentary to the high
angular resolution optical and UV observation  of the individual sources.

For the radio sources, these surveys operate at higher frequencies and are
complementary to low frequency compilations of radio point sources
(Condon et al.~\cite{condon98}, White et al.~\cite{white97}),
giving us the opportunity to study the spectral dependence of known sources
and to probe an entirely new population.
Since, in general, radio point sources have non-trivial spectra
(Herbig \& Readhead \cite{herbig92}),
the multi-frequency detection of such sources would be invaluable in our
understanding of the behavior of radio sources.
Under reasonable assumptions associating different populations of radio
sources (of a given lifetime) with halos
(Haiman \& Hui~\cite{haiman01}, Martini \& Weinberg~\cite{martini01}),
one would be able to constrain cosmological models through a study of
the clustering of radio sources.
At $1\,$mJy we expect about 1 source per 50 arcmin${}^2$, which would
allow APEX-SZ to constrain $1+w(\theta)$ on arcminute scales at the several
times $10^{-3}$ level.
Whether this is enough to perform a reliable measurement of $w(\theta)$
depends on the degree of bias and line-of-sight dilution of the clustering.
At lower frequencies, the angular correlation function
$w(\theta)\propto\theta^{1-\gamma}$ with $\gamma\simeq 1.8$.
The amplitude, $A$, of $w(\theta)$ depends on the flux limit and flattens
out for low flux limits.
Extrapolating from the NVSS data one would expect $A\lsim 10^{-3}$ at
$1\,$mJy (Overzier et al.~\cite{overzier03}), which would be near or
below the threshold for APEX-SZ.

Clearly to extract all of the excellent science that can be done with
these instruments will require coordinated observations over a range of
frequencies.  In this regard the sub-mm instruments intended for the
{\sl APEX\/} platform or the planned
SCUBA-2\footnote{http://www.roe.ac.uk/atc/projects/scuba\_two/}
with its higher angular resolution and frequency range will be particularly
valuable.  For the radio sources coordinated observations with the SZA or
other lower frequency instruments would be useful.

\section{Discussion and Conclusions} \label{sec:conclusions}
 
Advances in detector technology have brought us to the stage where
sensitive, high angular resolution, wide area surveys in the sub-mm
are practical.  Several new instruments are funded and under construction
which should map hundreds or thousands of square degrees of sky at arcminute
resolution with sensitivities of around $10\,\mu$K.  The stated aim of these
surveys is to find clusters of galaxies using the Sunyaev-Zel'dovich effect.

At these sensitivities, foregrounds are of particular importance, and in this
paper we have used new observational data on the IR and radio source counts
to estimate their impact on upcoming surveys. The contribution from both radio
and IR point sources are found to be non-negligible.
The IR point sources are of particular interest, because their number density
is higher than was expected based on early theoretical modeling.
Using a model of the source counts we predict that surveys with $1'$
resolution will be `confusion limited' before they reach $10\,\mu$K.
Typical effective noise levels from unsubtracted point sources can be more
than twice the nominal survey sensitivities at $150\,$GHz and $1'$ resolution
and even higher at $220\,$GHz.
The contamination by point sources has a strong impact on how such surveys
will study the thermal SZ effect, and may provide the ultimate limitation
to studies of the kinetic SZ effect and weak gravitational lensing.
Such surveys need to make multi-frequency observations or coordinate
observations with other instruments to allow them to remove point sources to
low flux levels.

Conversely these instruments will have a golden opportunity to study the
inverted spectrum radio sources underrepresented in current surveys, map
the FIB and probe the high-$z$ universe.
Not only will these upcoming surveys have the capability to detect thousands
of high redshift IR and radio sources to very low flux limits, they will
also be able to look into the correlations of the FIB and the clustering of
radio point sources.

\acknowledgments
M.~White thanks Bruce Partridge and Douglas Scott for helpful conversations.
We thank the Aspen Center for Physics, where this collaboration was begun,
for their support.
We also thank Gil Holder and Sarah Church for pointing out an error in
our treatment of radio sources in an earlier version of this paper.
This research was additionally supported by the NSF and NASA.


\begin{thebibliography}{99}

\bibitem[1998]{Barger}
Barger A.J., Cowie L.L., Sanders D.B., Taniguchi Y.
 1998, Nature, 394, 248 [astro-ph/9806317]
                                                                                
\bibitem[2003]{WMAP}
Bennett C.L., et al., 2003, \apj, in press [astro-ph/0302208]
                                                                                
\bibitem[1999]{Bir}
Birkinshaw M., 1999, Phys. Rep., 310, 98

\bibitem[1999]{borys99}
Borys, C., ,Chapman,, S.~C.,Scott, D., 1999, \mnras, 308, 527

\bibitem[2003]{borys03}
Borys, C., ,Chapman, S.~C., Halpern, M., Scott, D., 2003, [astro-ph/0301427]
    
\bibitem[1998]{condon98}
Condon, J. J. et al, 1998, \aj, 115, 1693 
                                                                                
\bibitem[1998]{Eales}
Eales S., Lilly S., Gear W., Dunne L., Bond J.R., Hammer F., Le Fevre O.,
 Crampton D. 1999, \apj, 515, 518 [astro-ph/9808040]
 
\bibitem[2000]{haiman00}
Haiman, Z., Knox, L., 2000, \apj, 530, 124

\bibitem[2001]{haiman01}
Haiman, Z., Hui, L., 2001,\apj, 547, 27

\bibitem[1992]{herbig92}
Herbig, T., Readhead, A. C. S., 1992, \apjs, 81, 83
 
\bibitem[2003]{hirata03} 
Hirata, C.~M., Seljak, U., 2003, \prd, 67, 43001

\bibitem[2002]{holder02}
Holder, G., 2002, \apj, 580, 36
                                                                                
\bibitem[1998]{Holletal}
Holland W.S., et al. 1998, Nature, 392, 788
                                                                                
\bibitem[1999]{Holland}
Holland W.S., et al. 1999, \mnras, 303, 659

\bibitem[2001]{hu01} 
Hu, W, 2001, \apjl, 557, L79

\bibitem[2003]{hu03b}
Hu, W., 2003,  \prd, 67, 081304
                                                                                
\bibitem[1998]{Hughes}
Hughes D.H., et al. 1998, Nature, 394, 241

\bibitem[2001]{knox01}
Knox, L., Cooray, A., Eisenstein, D., Haiman, Z., 2001, \apj, 550, 7

\bibitem[2002]{levin02}
Levine, E.~S., Schulz, A.~E., White, M., 2002, \apj, 577, 569

\bibitem[2001]{majumdar01}
Majumdar, S., 2001, \apjl, 555, L7

\bibitem[2003]{majumdar03}
Majumdar, S., Mohr, J.~J., 2003, \apj, 585, 603

\bibitem[2001]{martini01}
Martini, P., Weinberg, D. H., 2001, \apj, 547, 12

\bibitem[2003]{okamoto03} 
Okamoto, T., Hu, W.,2003, \prd, 67, 83002

\bibitem[2003]{overzier03}
Overzier R.A., R\"{o}ttgering H.J.A., Rengelink R.B., Wilman R.J., 2003,
  A\&A, 405, 530

\bibitem[1980]{LSSU}
Peebles P.J.E. 1980, The Large-Scale Structure of the University,
Princeton University Press, Princeton

\bibitem[2003]{PMBBZGSD}
Perrotta F., et al., 2003, \mnras, 338, 623

\bibitem[1996]{puget96}
Puget J.-L., et al. 1996, A\&A, 308, L5
                                                                                
\bibitem[1995]{Rep}
Rephaeli, Y., 1995, ARA\&A, 33, 541
                                                                                
\bibitem[1999]{ScoWhi}
Scott D., White M., 1999, A\&A, 346, 1 [astro-ph/9808003]

\bibitem[1999]{seljak99} 
Seljak, U., Zaldarriaga, M., 1999, Physical Review Letters, 82, 2636

\bibitem[1997]{SIB}
Smail I., Ivison R.J., Blain A.W. 1997, ApJ, 490, L5
                                                                                
\bibitem[1972]{SZ72}
Sunyaev R.A., Zel'dovich Ya. B., 1972, Comm. Astrophys. Space Phys., 4, 173
                                                                                
\bibitem[1980]{SZ80}
Sunyaev R.A., Zel'dovich Ya. B., 1980, ARA\&A, 18, 537

\bibitem[1996]{TegEfs}
Tegmark M., Efstathiou G. 1996, MNRAS, 281, 1297

\bibitem[2003]{Tru}
Trushkin S., 2003, Bull. Special Astrophys. Observatory, v.55, 90
[astro-ph/0307203]

\bibitem[2001]{weller02}
Weller, J., Battye, R., Kniessl, R., 2001, [astro-ph/0110353]

\bibitem[2002]{WHS}
White M., Hernquist L., Springel V., 2002, \apj, 579, 16 [astro-ph/0205437]

\bibitem[1997]{white97}
White, R. L. et al., 1997, \apj, 475, 479

\bibitem[1999]{zaldarriaga99} 
Zaldarriaga,M., Seljak, U., 1999, \prd, 59, 123507

\bibitem[2002]{zhang02}
Zhang, P., Pen, U., Wang, B., 2002, \apj, 577, 555

\bibitem[2003]{zhang03}
Zhang, P., Pen, U., Trac, H., 2003, [astro-ph/0304534]                                                                               
\end{thebibliography}
\end{document}